\documentclass{PoS}
\usepackage{amsmath}

\title{B meson decays from moving-NRQCD on fine MILC lattices}
\ShortTitle{B meson decays from moving-NRQCD on fine MILC lattices}

\author{C.~T.~H.~Davies and \speaker{K.~Y.~Wong}\\
        Department of Physics and Astronomy,
        University of Glasgow, Glasgow, G12 8QQ, United Kingdom}

\author{G.~P.~Lepage\\
        Laboratory of Elementary Particle Physics,
        Cornell University, Ithaca, New York 14853, USA}

\abstract{Lattice simulations of $B\to\pi l\nu$ decays are
problematic in the low $q^{2}$ region when the pion has to
have large momentum in the lattice frame for a $B$ meson
at rest or moving slowly. The moving-NRQCD formalism provides
a way around this by giving an accurate discretization of
a $b$ quark moving at arbitrary velocity $v$, and therefore
a $B$ meson moving at high momentum in the lattice frame.
Here we show results from simulations of the moving-NRQCD
action complete through $O(1/M)$ coupled to the asqtad action
on a MILC fine ensemble. We show $B$, $B^{*}$ masses
as a function of $v$ and demonstrate how
to determine the meson kinetic mass and the renormalization
of $v$ non-perturbatively. We will also discuss the perturbative
renormalization of the moving-NRQCD action from high-$\beta$
Monte Carlo simulations.}

\FullConference{XXIVth International Symposium on Lattice Field Theory\\
                July 23-28, 2006\\
                Tucson, Arizona, USA}

\begin{document}

\section{INTRODUCTION}

Precise determination of the Cabibbo-Kobayashi-Maskawa (CKM)
matrix is crucial for high precision tests of the standard
model. The matrix element $|V_{ub}|$ can be determined
from studies of $B\to\pi l\nu$ semileptonic decays in
B-factories. Such decays, however, involve strong interaction
dynamics also in the form of hadronic matrix elements; therefore
precise determination of $|V_{ub}|$ requires accurate calculation
of these form factors.

\begin{figure}
\centering
\includegraphics[width=0.45\textwidth]{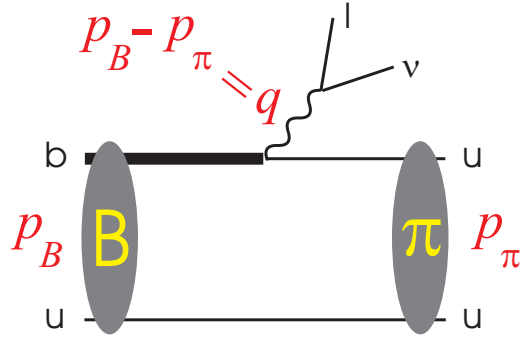}
\caption{Kinematics of $B\to\pi l\nu$ semileptonic decay.
$p_{B}$ is the momentum of the $B$ meson and $p_{\pi}$ is the
momentum of the pion, $q$ is the momentum transfer.}
\label{fig_btopidiag}
\end{figure}

Lattice QCD provides a first principles non-perturbative
approach for calculating the form factors $f_{+}(q^{2})$,
$f_{0}(q^{2})$ in $B\to\pi l\nu$ decays. Fig.~\ref{fig_btopidiag}
shows the kinematics of the process. Lattice simulations
are problematic, however, when the pion has very large
recoil momentum, i.e., when $q^{2}$ is small. This is because
discretization errors increase as the pion momentum increases;
and statistical errors become worse when the hadrons have large
momenta. Lattice results are therefore available at the large
$q^{2}$ region only~\cite{Gulez06}. These simulations
were done in the $B$ meson rest frame, and two different
techniques were used to handle the $b$ quark on the lattice:
non-relativistic QCD (NRQCD) and the Fermilab formalism.
Because of the lack of data at small $q^{2}$ error in $|V_{ub}|$
is dominated by error in lattice results.

This problem can be solved in two stages. First instead of
working in the $B$ meson rest frame we choose a moving frame in
which the $B$ meson is moving in the opposite direction of the pion.
This reduces the momentum of the pion, but at the expense of
increasing finite-lattice-spacing errors for the now high-momentum
$B$ meson. The increased errors in the $B$ system are
mostly associated with the $b$ quark since most of its
momentum is carried by the heavy quark. We work around this
problem by using the moving-NRQCD (mNRQCD)
formalism~\cite{Sloan98,KMFThesis}
for the $b$ quark.

Moving-NRQCD provides an accurate discretization of a heavy quark
moving at arbitrary velocity $v$. It is useful to
parameterize the $b$ quark momentum as $p_{b}=um_{b}+k$ where
$u=p_{B}/m_{B}=(E_{B}/m_{B},v)$ is the
4-velocity of the $B$ meson. The residual momentum $k$ is much smaller
than $p_{B}$ and $k^{2}\sim O(\Lambda_{QCD}^{2})$.
In mNRQCD one treats the velocity $v$ exactly and discretizes
the residual momentum only. The leading discretization errors
therefore become $O(a^{2}k^{2})$ instead of
$O(a^{2}p_{B}^{2})$, which could be as large as several
$\mathrm{GeV}^{2}$. We refer the readers to Refs.~\cite{Sloan98,KMFThesis}
for details of the formulation.

The mNRQCD action can be derived using the FWT transformation
or by boosting the NRQCD action to a moving frame~\cite{Sloan98,KMFThesis}.
In this project we use an action which is complete through
$O(1/M)$ and tadpole improved (the tadpole factor $u_{0}$ is
implicit in the equations):
\begin{equation}
\begin{array}{l}
\textrm{\underline{Evolution equation:}}  \vspace{0.3cm} \\
\qquad \qquad
{\displaystyle
G_{x,t+1}=
\left(1-\frac{\delta H}{2}\right)
\left(1-\frac{\delta H_{0}}{2n}\right)^{n}
U_{x,\hat{t}}^{\dag}
\left(1-\frac{\delta H_{0}}{2n}\right)^{n}
\left(1-\frac{\delta H}{2}\right)
G_{x,t}}                                  \vspace{0.3cm} \\
\mathrm{\underline{Action:}}              \vspace{0.3cm} \\
\qquad \qquad
{\displaystyle
H_{0}=-iv\cdot\Delta^{\pm}
-\frac{\Delta^{(2)}}{2M_{0}\gamma}}       \vspace{0.3cm} \\
\qquad \qquad
{\displaystyle
\delta H=
-\frac{1}{2M_{0}\gamma}\sigma\cdot\tilde{B}_{m}
+\left(\frac{1}{2M_{0}\gamma}+\frac{1}{4n}\right)
\left(v\cdot\Delta^{\pm}\right)^{2}
-\frac{i}{4M_{0}n\gamma}\left(v\cdot\Delta^{\pm}\right)
\Delta^{(2)}}                             \vspace{0.2cm} \\
\qquad \qquad \hspace{1cm}
{\displaystyle
+\frac{i}{6}\sum_{i}v_{i}\Delta_{i}^{+}\Delta_{i}^{\pm}\Delta_{i}^{-}
+\frac{1}{24M_{0}\gamma}\sum_{i}\Delta_{i}^{(4)}
-\frac{1}{6M_{0}\gamma}
\left(v\cdot\Delta^{\pm}\right)
\sum_{i}v_{i}\Delta_{i}^{+}\Delta_{i}^{\pm}\Delta_{i}^{-}} \vspace{0.3cm} \\
\qquad \qquad
{\displaystyle
\tilde{B}_{m}=\gamma
\left(\tilde{B}-v\times\tilde{E}
-\frac{\gamma}{\gamma+1}v
\left(v\cdot\tilde{B}\right)\right)}
\end{array}
\label{eq_mnrqcdaction}
\end{equation}
where $\gamma=1/\sqrt{1-v^{2}}$, $M_{0}$ is the input bare mass,
$n$ is the stability parameter, and $\tilde{B}_{m}$ is the
magnetic field in the moving frame (we use $O(a^{2})$ improved electric field
$\tilde{E}$ and magnetic field $\tilde{B}$).
The corresponding dispersion relation is given by
\begin{equation}
E(k)
=E_{0}+\sqrt{P_{tot}^{2}+M_{kin}^{2}}
=E_{0}+\sqrt{(Z_{p}P_{0}+k)^{2}+Z_{m}^{2}M_{0}^{2}}
\label{eq_disp}
\end{equation}
where $P_{0}=\gamma M_{0}v$. Here $E_{0}$ is
the zero energy, $Z_{m}$ is the mass renormalization and $Z_{p}$ is
the external momentum renormalization, which is a new feature in mNRQCD.
These renormalization constants have perturbative expansions
\begin{eqnarray}
E_{0}(v) & = & c_{1,E0}(v)\alpha_{s}+\ldots, \\
Z_{m}(v) & = & 1+c_{1,Zm}(v)\alpha_{s}+\ldots, \\
Z_{p}(v) & = & 1+c_{1,Zp}(v)\alpha_{s}+\ldots
\end{eqnarray}
Notice the dependence on the velocity. In the next section
we will discuss how to extract the perturbative coefficients $c_{1}$,
$c_{2}$ \ldots from high-$\beta$ Monte Carlo simulations.

\section{PERTURBATIVE EXPANSIONS FROM HIGH-$\beta$ (WEAK COUPLING)
SIMULATIONS}

Knowing how the bare action parameters renormalize on the lattice
is crucial for accurate simulations. The first order coefficients
(quenched with the Wilson plaquette action, $\delta H=0$) have
been computed in standard lattice perturbation theory using diagrammatic
techniques~\cite{Dougall05}. Analytic calculations, however,
are very challenging because of the proliferation of diagrams
due to the lattice cutoff. This is particularly true at higher orders.
An alternative is to use numerical methods. Here we determine
the first order coefficients from high-$\beta$ simulations~\cite{Dimm95}
and compare the results to the analytic calculations.

We start by doing ten sets of simulations, at volumes
$V=L^{3}\times T=6^{3}\times 12$, $(8^{3},10^{3},12^{3})\times 16$,
with $\beta$-values range from $10.0$ to $60.0$. The corresponding
$\alpha_{s}$ are of the order of $0.01$-$0.1$. It is important to work
at small couplings so that the theory enters the perturbative phase.
Twisted boundary conditions are also used in the $x$ and $y$ directions
to suppress non-perturbative effects due to Z(3) phases~\cite{Dimm95}.
In each simulation we measure the heavy quark energy at several
momenta and fit the results to Eq.~\ref{eq_disp} to extract the
renormalization constants. We were able to work at small momentum,
$k^{2}<0.3$, by using generalized boundary conditions for the
quark field in the $z$ direction~\cite{Divitiis04}. Finally to
obtain the perturbative coefficients we fit the renormalization
constants obtained at different $\beta$ to a polynomial
in $\alpha_{s}$.

\begin{figure}
\centering
\includegraphics[width=0.5\textwidth]{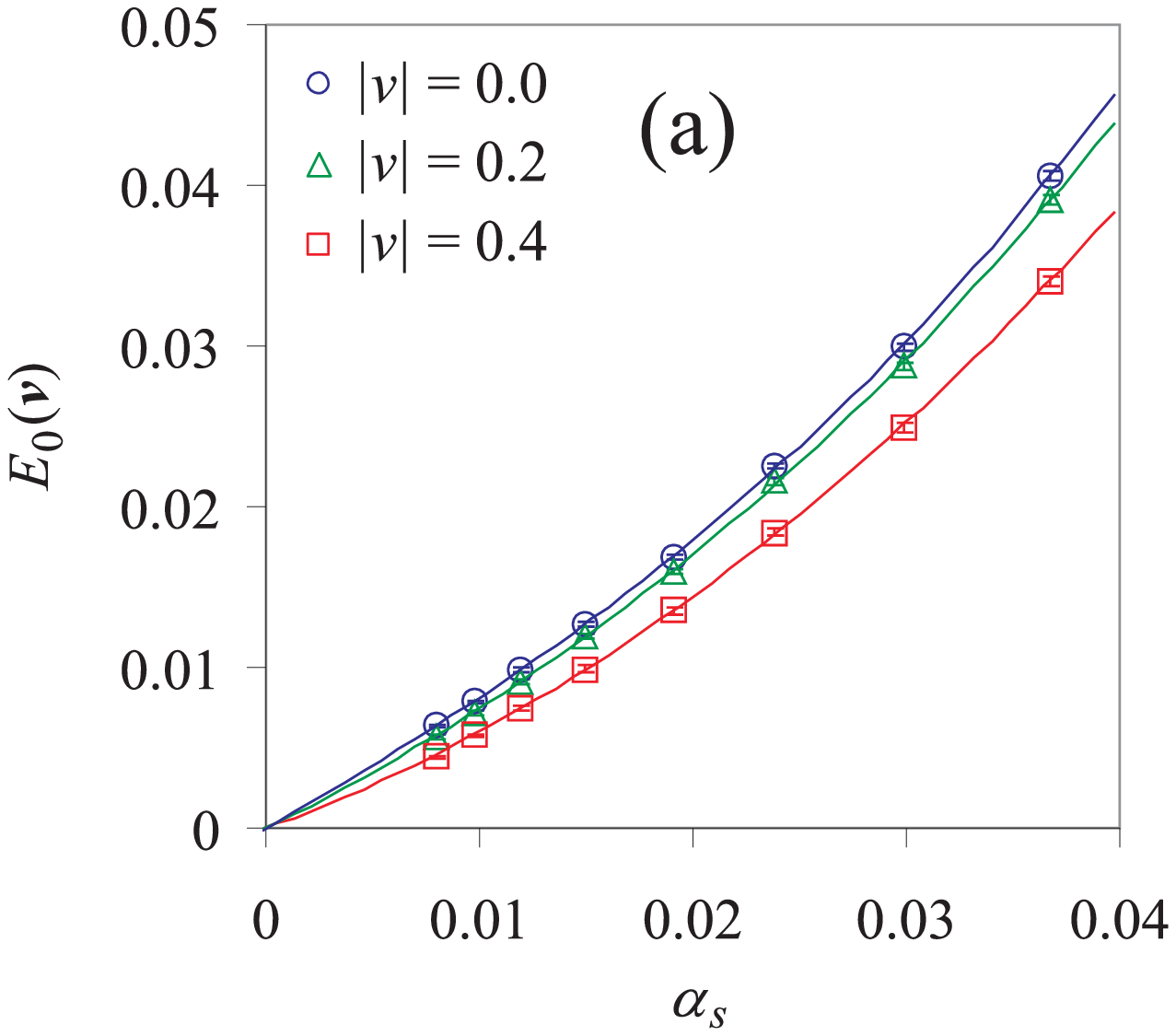}%
\includegraphics[width=0.5\textwidth]{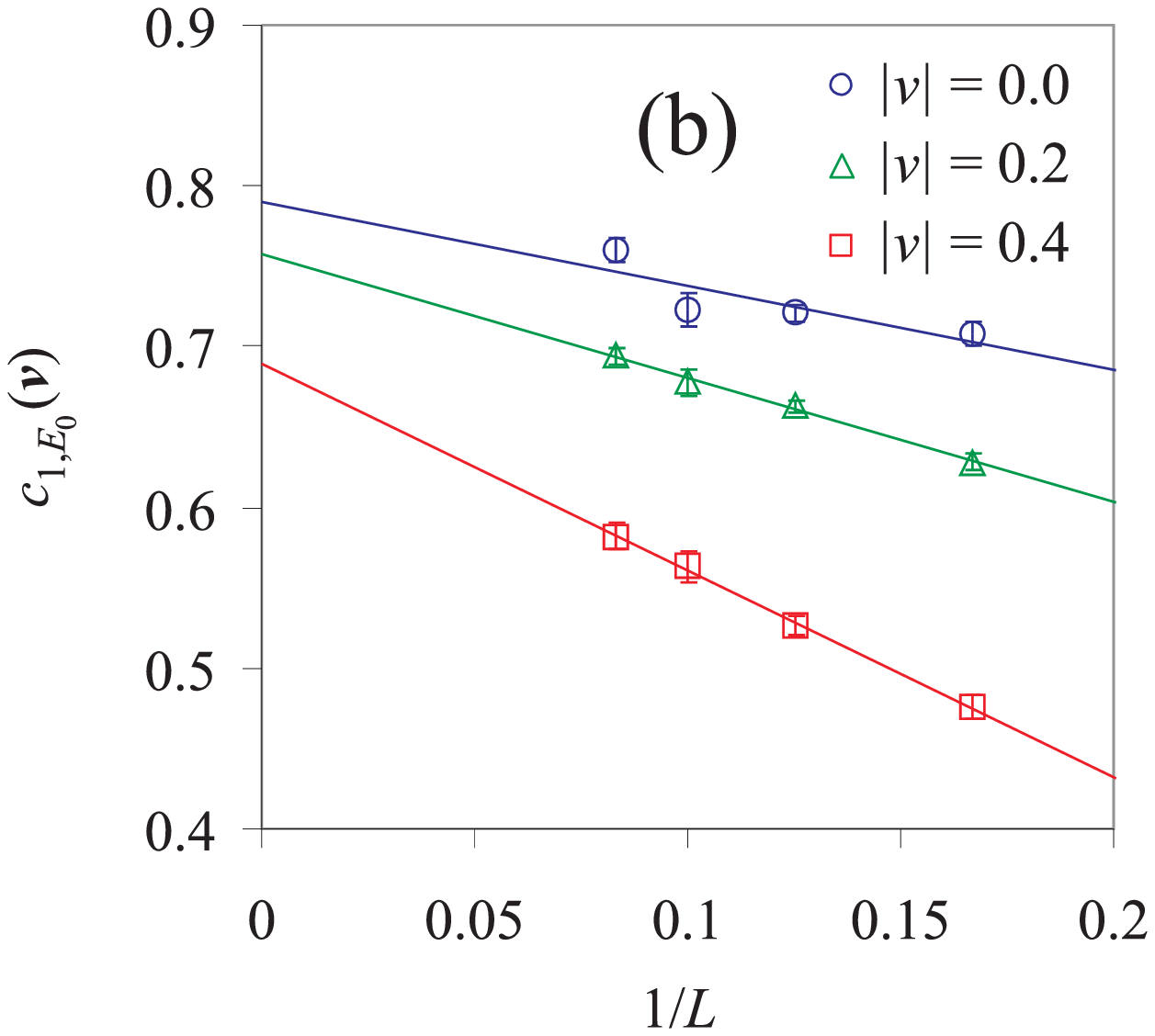}
\caption{(a) Zero energy $E_{0}(v)$ as a function of
$\alpha_{s}$. The volume is $6^{3}\times 12$. (b) $1/L$ infinite volume
extrapolation of the first order coefficients $c_{1,E0}(v)$.}
\label{fig_E0highbeta}
\end{figure}

\begin{figure}
\centering
\includegraphics[width=0.5\textwidth]{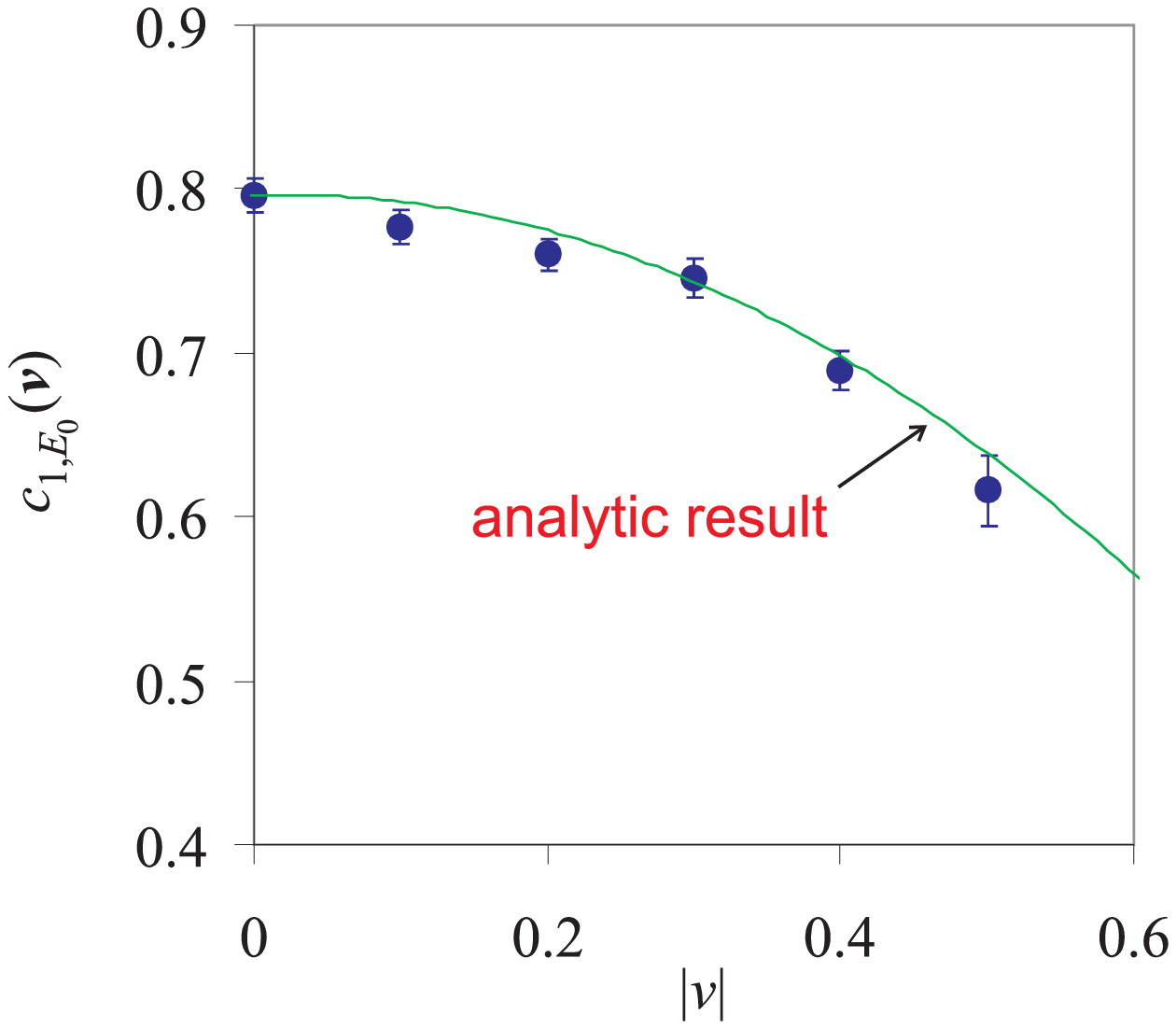}
\caption{Zero energy first order perturbative coefficients
$c_{1,E0}(v)$ after infinite volume extrapolation.
The curve shows the analytic results obtained in Ref.~\cite{Dougall05}.}
\label{fig_E0c1result}
\end{figure}

Fig.~\ref{fig_E0highbeta}a plots the zero energy $E_{0}(v)$
against $\alpha_{s}$ at several velocities at volume $6^{3}\times 12$.
The ``slopes'' of the lines are the first order coefficients
$c_{1,E0}(v)$ and the ``curvatures'' are the second order
coefficients. Fig.~\ref{fig_E0highbeta}b shows the $1/L$ infinite volume
extrapolation of the first order coefficients $c_{1,E0}(v)$;
and the final results are given in Fig.~\ref{fig_E0c1result}.
Our numbers agree very well with the analytic results
(given by the curve in the graph). We have also checked that
reversing the fit order, i.e., extrapolate to infinite volume at each
$\beta$ and then fit to an expansion in $\alpha_{s}$, gives consistent
results.

\begin{figure}
\centering
\includegraphics[width=0.5\textwidth]{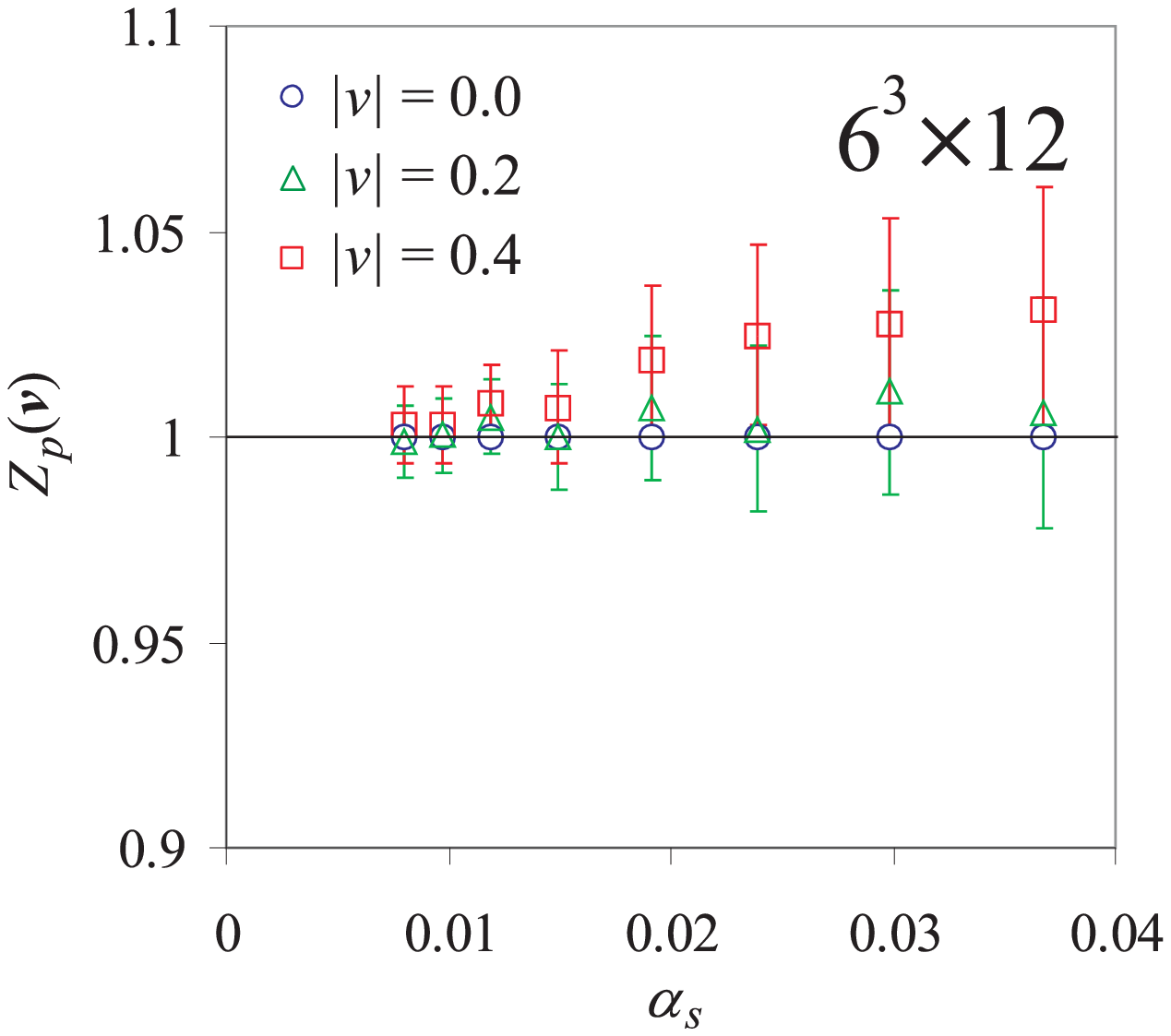}%
\includegraphics[width=0.5\textwidth]{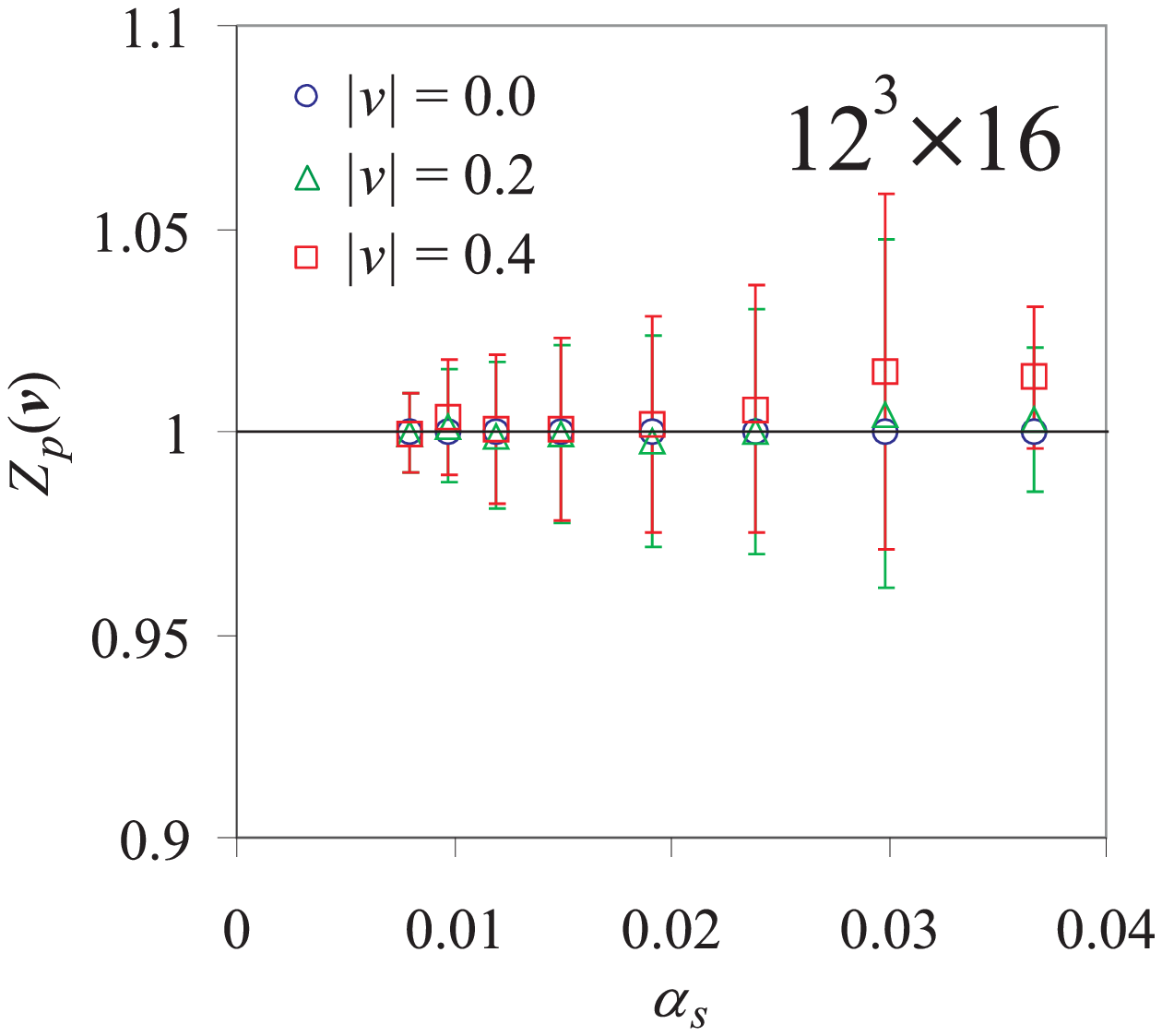}
\caption{External momentum renormalization $Z_{p}(v)$
as a function of $\alpha_{s}$. Results are shown for volumes
$6^{3}\times 12$ and $12^{3}\times 16$.}
\label{fig_vrhighbeta}
\end{figure}

Fig.~\ref{fig_vrhighbeta} shows the external momentum renormalization
$Z_{p}(v)$ obtained at $6^{3}\times 12$ and
$12^{3}\times 16$. Despite the large statistical errors all the data
are consistent with 1. We can conclude that there is
little or no renormalization of the bare external momentum. This is consistent
with the fact that $v$ is protected from renormalization
because of approximate re-parameterization invariance of the
action~\cite{Sloan98,KMFThesis}.

\section{$B$ MESON DECAYS ON MILC FINE LATTICES}

In this section we present results for $B$ meson correlators on MILC
fine lattices. We study the dependence of $B$ and $B^{*}$ masses
on $v$. We also show how to extract the momentum
renormalization from simulations.

Simulations were done with the $O(1/M)$ accurate tadpole improved
($u_{0}=0.8541$) mNRQCD action at three different velocities
$v=(v_{x},0,0)$, $v_{x}=0.0$, $0.25$ and $0.5$.
The MILC fine lattices have a lattice spacing of $2.258\mathrm{GeV}^{-1}$,
which is set by the $\Upsilon$ 2S-1S mass splitting~\cite{Bernard01};
the sea quark mass is $u_{0}am_{f_{1,2}}=0.0062,0.031$.
The bare mass is set to $aM_{0}=1.95$ ($n=2$) for the heavy quark;
and we use the asqtad formalism for valence light quarks
with a quark mass $am_{q}=0.031$. It has been shown that, from
previous NRQCD studies on heavy-light systems and the
$\Upsilon$ spectrum, these quark masses reproduce the correct
$B_{s}$ meson mass at $v=0$. Gaussian smearings with a
radius of $2a$ were applied to the heavy quark at both the source and
the sink. Once again we measure the meson energy at different
momenta, and fit the results to Eq.~\ref{eq_disp} to extract the
zero energy $E_{0}(v)$, kinetic mass
$m_{kin}(v)$, and momentum renormalization
$Z_{p}(v)$. Thirty nine different momenta, with
$0\leq k^{2}\leq 0.45$, were used.

\begin{figure}
\centering
\includegraphics[width=0.5\textwidth]{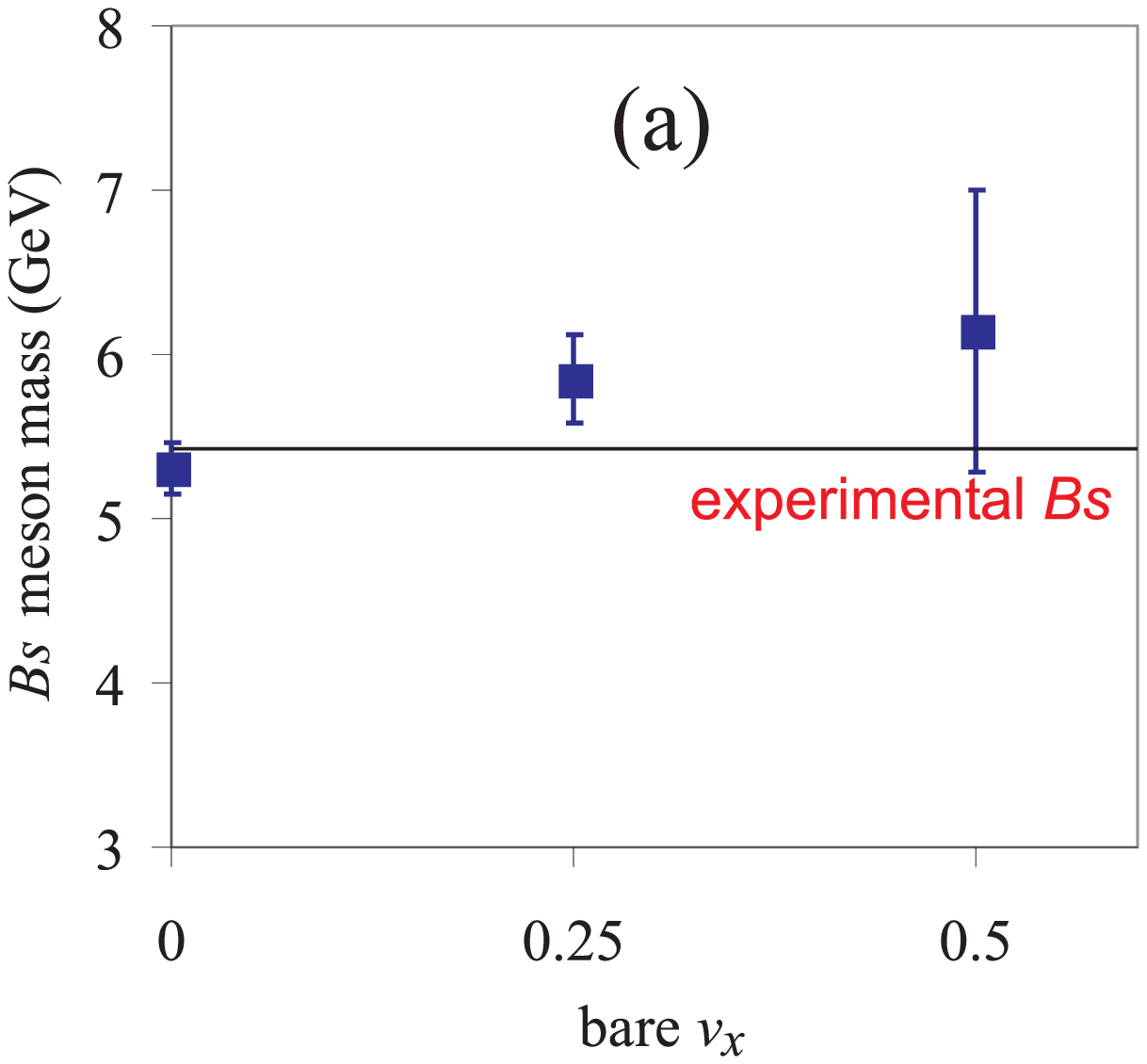}%
\includegraphics[width=0.5\textwidth]{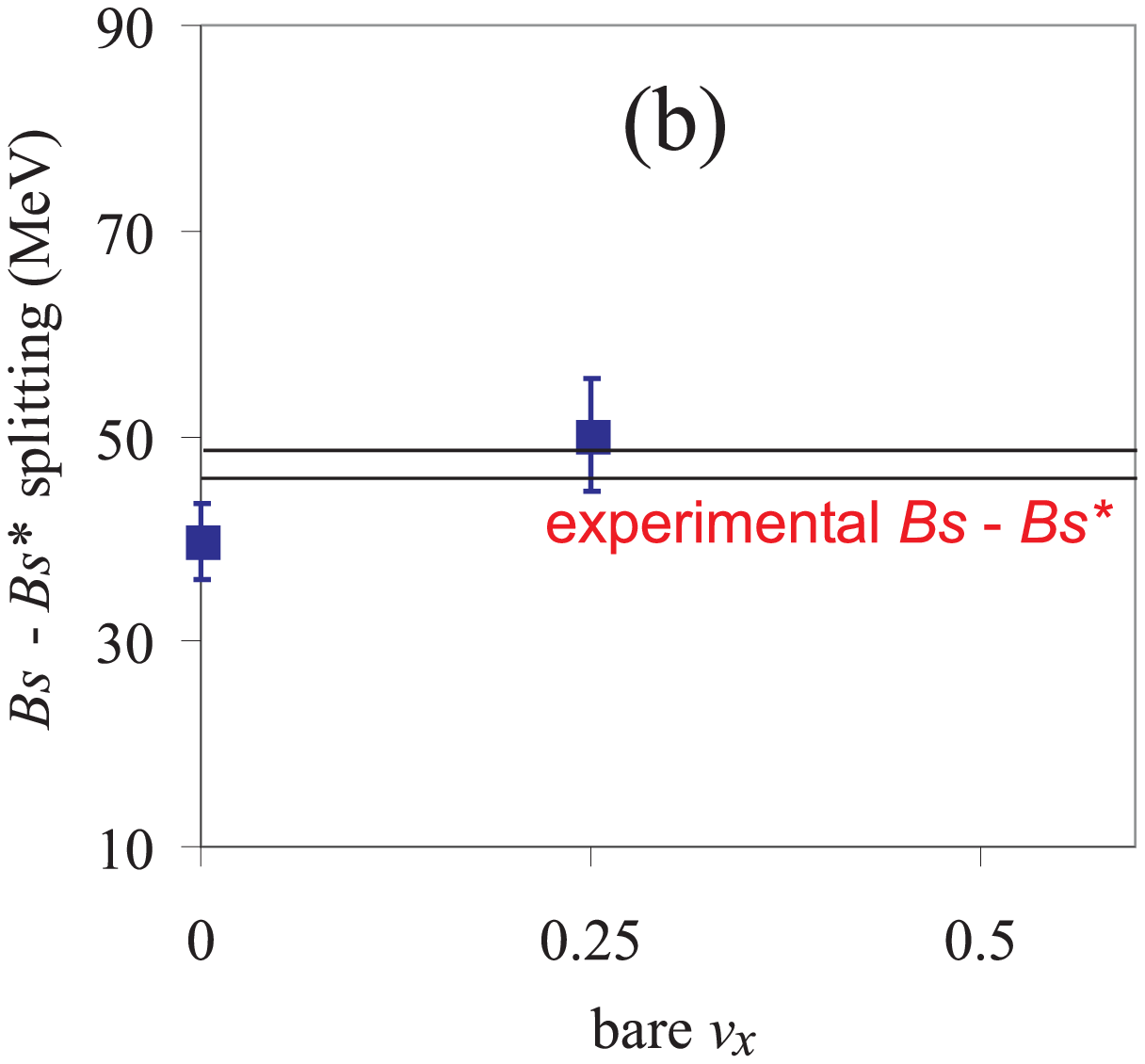}
\caption{(a) $B_{s}$ meson mass $m_{kin}$ as a function of bare velocity.
(b) $B_{s}$-$B_{s}^{*}$ energy splitting as a
function of $v$. The experimental values are given
by the horizontal lines.}
\label{fig_Bsmass}
\end{figure}

Fig.~\ref{fig_Bsmass}a plots the $B_{s}$ mass as a function of
bare velocity $v_{x}$. Result at $v=0$
agrees well with the experimental value as suggested in the last
paragraph. Our data show almost no dependence on the velocity,
although a small increase can be observed as $v_{x}$
becomes large. This suggests that one may need to tune (reduce)
the input bare mass $M_{0}$ as $v$ increases.

The energy splitting of $B_{s}$-$B_{s}^{*}$ is plotted in
Fig.~\ref{fig_Bsmass}b. Again results show only a small velocity
dependence. Data at $v=0$, however,
is about one $\sigma$ lower than the experimental value.
It is believed that radiative correction of the
$\sigma\cdot B$ term in the action is
responsible for this deviation from the experimental data.

\begin{figure}
\centering
\includegraphics[width=0.5\textwidth]{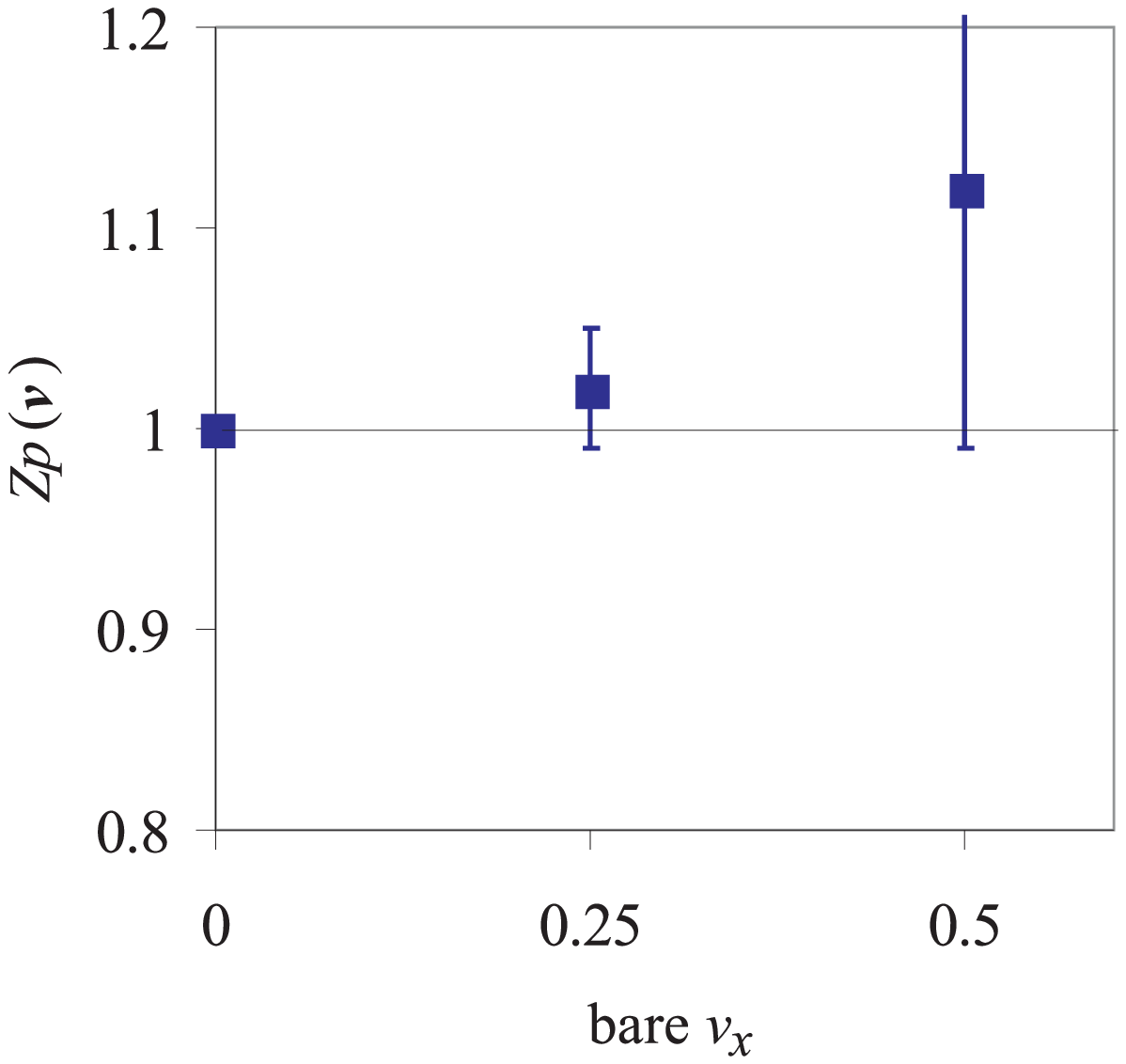}
\caption{Momentum renormalization $Z_{p}(v)$ of
$B_{s}$ as a function of input velocity.}
\label{fig_Zv}
\end{figure}

Finally the external momentum renormalization $Z_{p}(v)$
for $B_{s}$ meson is
given in Fig.~\ref{fig_Zv}. All the data points are consistent
with 1. This agrees with the results obtained in perturbation
theory (see Fig.~\ref{fig_vrhighbeta} in the last section), and
again reflects the re-parameterization invariance
property of the action.

Fig.~\ref{fig_Bsmass} and \ref{fig_Zv} both reveal that
statistical errors increase significantly between $v_{x}=0.25$
and $0.5$. It is therefore advisable to use a more advanced
smearing technique, such as the so-called ``magic''
smearing~\cite{KMFThesis},
in future simulations.

\section{CONCLUSION}

Moving-NRQCD provides a promising approach for studying
semileptonic decays at high recoil on the lattice.
This project represents a major step forward towards this
goal. We have demonstrated that perturbative expansions can be
computed in high-$\beta$ simulations with much less effort.
First order results agree very well with the results obtained
using analytic methods. Work is currently underway to
extract higher-order coefficients from the data. In the
second part of the paper we present results for $B$ meson
correlators on MILC fine lattices. A variety of quantities have been
measured at different $B$ meson velocities; and we observe little
or no velocity dependence. This
is significant since it indicates that we do not need to
tune the input bare parameters as $v$ changes.
On the other hand our results show a sharp increase in
statistical errors as $v$ becomes large.
This suggests that it may be difficult to reach a very low
$q^{2}$ value. It is, however, not necessary to give the form
factors at very small $q^{2}$ since the main experimental
focus is on $8\mathrm{GeV}^{2}\leq q^{2}\leq 16\mathrm{GeV}^{2}$.
Simulations for $B\to\pi l\nu$ decays will start soon and
we expect to have new results in the next year.

\section{ACKNOWLEDGMENTS}

This work was supported by PPARC (UK) and the DOE and
NFS (USA). We thank the MILC Collaboration for making their
unquenched gauge configurations available and the
Fermilab Collaboration for use of their light propagators
on the fine lattices. We thank Alex Dougall, Kerryann Foley
and Junko Shigemitsu for useful conversations.


\begin{thebibliography}{99}

\bibitem{Gulez06}
E.~Gulez {\it et al.}, Phys.~Rev.~D~{\bf 73}, 074502 (2006);
M.~Okamoto, Proc.~Sci., LAT2005 (2005) 013.

\bibitem{Sloan98}
J.~H.~Sloan, Nucl.~Phys.~Proc.~Suppl.~{\bf 63}, 365 (1998);
K.~M.~Foley and G.~P.~Lepage, Nucl.~Phys.~Proc.~Suppl.~{\bf 119}, 635 (2002);
A.~Dougall {\it et al.}, LAT2005 (2005) 219.

\bibitem{KMFThesis}
K.~M.~Foley, Ph.D. Thesis, Cornell University (2004).

\bibitem{Dougall05}
A.~Dougall {\it et al.}, Nucl.~Phys.~Proc.~Suppl.~{\bf 140}, 431 (2005).

\bibitem{Dimm95}
W.~Dimm {\it et al.}, Nucl.~Phys.~Proc.~Suppl.~{\bf 42}, 403 (1995);
K.~J.~Juge, Nucl.~Phys.~Proc.~Suppl.~{\bf 94}, 584 (2001);
H.~D.~Trottier {\it et al.}, Phys.~Rev.~D~{\bf 65}, 094502 (2002);
K.~Y.~Wong, H.~D.~Trottier and R.~M.~Woloshyn,
Phys.~Rev.~D~{\bf 73}, 094512 (2006).

\bibitem{Divitiis04}
G.~M.~de~Divittis, R.~Petronzio and N.~Tantalo, Phys.~Lett.~B~{\bf 595},
408 (2004);
J.~M.~Flynn {\it et al.}, Phys.~Lett.~B~{\bf 632}, 313 (2006).

\bibitem{Bernard01}
C.~Bernard {\it et al.}, Phys.~Rev.~D~{\bf 64}, 054506 (2001);
A.~Gray {\it et al.}, Phys.~Rev.~D~{\bf 64}, 094507 (2005).

\end{thebibliography}
\end{document}